\documentclass[notitlepage,floats,aps,nofootinbib,preprintnumbers,superscriptaddress,twocolumn,prl]{revtex4-1}

\usepackage{amsmath,amssymb,amsfonts}
\usepackage{graphicx}
\usepackage{xcolor}
\usepackage{mathrsfs}
\usepackage{slashed}
\usepackage[utf8x]{inputenc}
\usepackage{hyperref}
\hypersetup{colorlinks,citecolor=blue}

\newcommand{\BH}{\mathrm{BH}}
\newcommand{\PBH}{\mathrm{PBH}}
\newcommand{\DM}{\mathrm{DM}}
\newcommand{\FFP}{\mathrm{FFP}}

\def\beq{\begin{equation}\begin{aligned}}
\def\eeq{\end{aligned}\end{equation}}

\begin{document}
\preprint{IPPP/19/73}
\title{ \vspace*{-4mm} What if Planet 9 is a Primordial Black Hole? \vspace*{-2mm} }

\author{Jakub Scholtz}
\affiliation{Institute for Particle Physics Phenomenology, Durham University, Durham, DH1 3LE, United Kingdom}
\author{James Unwin}
\affiliation{Department of Physics, University of Illinois at Chicago, Chicago, IL 60607, USA;\\
\& Department of Physics, University of California, Berkeley \& Theoretical Physics Group, LBNL \& Mathematics Sciences Research Institute, Berkeley, CA 94720, USA}

\begin{abstract} 
We highlight that the anomalous orbits of Trans-Neptunian Objects (TNOs) and an excess in microlensing events in the 5-year OGLE dataset can be simultaneously explained by a new population of astrophysical bodies with mass several times that of Earth ($M_\oplus$). We take these objects to be primordial black holes (PBHs) and point out the orbits of TNOs would be altered if one of these PBHs was captured by the Solar System, inline with the Planet 9 hypothesis. Capture of a free floating planet is  a leading explanation for the origin of Planet 9 and we show that the probability of capturing a PBH instead is comparable. The observational constraints on a PBH in the outer Solar System significantly differ from the case of a new ninth planet. This scenario could be confirmed through annihilation signals from the dark matter microhalo around the PBH.

\end{abstract}

\maketitle

{\bf 1.~Introduction.}
As of this year, two gravitational anomalies of similar mass but very different origins remain to be explained. First, there is a growing body of observational anomalies connected to the orbits of trans-Neptunian Objects (TNOs) \cite{Brown04,Trujillo,Batygin}. These observations have been taken as evidence of a new ninth planet in our solar system, called Planet 9 (P9),  with mass $M_9\sim5-15M_\oplus$ and orbiting around the Sun at a distance of $300-1000\;\mathrm{AU}$ \cite{Batygin-PR}. Second, there is set of gravitational anomalies recently observed by the Optical Gravitational Lensing Experiment (OGLE). OGLE reported an excess of six ultrashort microlensing events with 
crossing times of $0.1 -0.3$ days \cite{Mroz}. The lensing objects are located towards the galactic bulge, roughly 8kpc away. These events correspond to lensing by objects of mass $M\sim 0.5M_\oplus - 20 M_\oplus$ \cite{Niikura:2019kqi} and could be interpreted as an unexpected population of free floating planets (FFPs) or as Primordial Black Holes (PBHs). 

It is remarkable that these two anomalies correspond to a similar mass scale. Perhaps the most natural explanation is that they are caused by the existence of an unknown population of planets, i.e.~the OGLE anomaly should be interpreted as due to FFPs denser than the local star population \cite{OGLE2} and P9 might be one of those planets that have been captured by the Solar System. This would imply that our models for planet formation may need to be updated to account for this new population of FFPs, but the current program to hunt for P9 would go unchanged.

We focus on a more exciting possibility: if the OGLE events are due to a population of PBHs then it is possible that the orbital anomalies of TNOs are also due to one of these PBHs that was captured by the Solar System. In this paper, we argue that this scenario is not unreasonable and discuss the observable implications; we estimate the probability of capture of a PBH by the Solar System, highlight that the observational constraints differ significantly between planets and PBHs, and point out that the dark matter (DM) microhalo, which generically forms around such a PBH, can lead to its discovery.


{\bf 2.~Two Anomalies.} 
 While the structure of the Solar System to semi-major axis $a\sim100$ AU is well explained, for $a>250$ AU there are TNO populations whose orbits cannot be readily understood. Observations of TNOs, objects with $a>30$ AU, and extreme TNOs (eTNOs) with $a> 250$ AU exhibit the following anomalies:
\begin{itemize}

\item[i)] Unexpected clustering in eTNO orbits  \cite{Trujillo,Batygin};

\vspace{-1mm}

\item[ii)] The existence of high perihelia ($q\sim70$ AU) TNOs, such as Sedna,  collectively called Sednoids \cite{Trujillo,Sheppard,Brown04};

\vspace{-1mm}

\item[iii)] TNOs moving roughly perpendicularly  to the planetary plane (with inclination $i\gtrsim50^{\circ}$) \cite{Chen,Gomes,Gladman}. 

\end{itemize}
An excellent review of TNO anomalies is given in \cite{Batygin-PR}.

Simulations and analytic arguments indicate that these observations are at odds with the predicted dynamics assuming only the known giant planets. For instance, any coincidence of initial orbits of eTNOs will disperse under evolution (on a timescale of a 10-100 million years \cite{Brown}), and Solar System simulations imply inclinations are typically bounded by $i\lesssim40^{\circ}$ \cite{Levison,Nesvorny}.

Notably, all of the TNO anomalies can be simultaneously explained by a new gravitational source in the outer Solar System. From observations of TNO dynamics one can infer the likely mass and orbit of this hypothetical source, referred to as Planet 9. Simulations of TNOs have identified a number of characteristic benchmark scenarios for a new gravitational  source of mass 5$M_\oplus$ and 10$M_\oplus$, which we summarise in Table \ref{table1}. 

\begin{table}[h]
{\renewcommand{\arraystretch}{1.4}
\begin{tabular}{c|cccccc}
Benchmark & $~a$ (AU) ~& ~$e$~ & ~$i$ (deg)~ &~ $r_p$  (AU)~ &~ $r_a$ (AU) \\\hline
$5M_\oplus$   &  450  & 0.2 & 20  & 400 & 500  \\
$10M_\oplus$   &700  & 0.4& 15  & 420  & 980 
\end{tabular}
}
\caption{This table contains some benchmark P9 scenarios from \cite{Batygin-PR} stating central values. Here $a$ is the central major axis, $e$ is the ellipticity, $i$ is the inclination. Also given are the perihelion $r_p=r_a(1-e)$ and the aphelion $r_a=2ra-r_p$. Deviations of $\mathcal{O}(10\%)$ in parameters provide comparable fits. \label{table1}}
\end{table}

The second set of anomalies has been observed by OGLE. The OGLE observations, when interpreted as a population of PBHs, are consistent with a range of masses and density fractions \cite{Niikura:2019kqi} with
\beq
M\in [0.5 M_\oplus, 20 M_\oplus], & & f_\PBH \in [0.005,0.1],
\eeq
where $f_{\rm PBH}\equiv \Omega_{\rm PBH}/ \Omega_{\rm DM}$, in terms of the DM relic density $\Omega_{\rm DM}h^2\approx0.1$ with the PBH population forming a subcomponent with relic density $\Omega_\PBH$. The masses and density fractions are correlated with larger masses corresponding to smaller $f_{\rm PBH}$.

While an $M_\oplus$ object is too light to be an astrophysical black hole formed by stellar collapse, PBHs arise from over densities in the early universe  \cite{Hawking:1971ei,Carr:1974nx} and as a result can be substantially lighter than $M_\odot$. Formation of PBHs inferred from OGLE has been discussed in \cite{Tada:2019amh,Fu:2019ttf,Carr:2019kxo}.
We note that PBHs arise from $\mathcal{O}(1)$ density fluctuations during radiation domination, due to an increase  in the primordial power spectrum. An intriguing coincidence is that PBHs formed during radiation domination via a strong first order phase transition around the electroweak scale are expected to have mass of the same order as P9 with $M_{\rm BH}\sim M_\oplus(125~{\rm GeV}/T)^2$ \cite{Jedamzik:1996mr}.

Since \cite{Niikura:2019kqi} prefers $f_{\rm PBH}<1$, we assume DM particles account for the remaining fraction: $f_{\rm DM}=1-f_{\rm PBH}\approx 1$.
In that case, a PBH will accrete DM and form a microhalo \cite{Eroshenko,Boucenna:2017ghj,Bertone:2005xz,Adamek:2019gns,Lacki:2010zf}.  Since the DM densities in these microhalos are typically very high, DM annihilations can be significantly enhanced, leading to potentially detectable signals as we will discuss in Sections 4 and 5.


\vspace{3mm} 
{\bf 3.~Capture Probability.}
There are three alternative hypotheses for the origin of P9: a) P9 formed on its current orbit (`in situ'); b) P9 formed in the inner Solar System and has been up-scattered into its current orbit; or c) P9 has formed outside of the Solar System and has been captured. While all three scenarios are unlikely, they are still favourable compared to the chance alignment of TNOs \cite{Batygin}. In case of the in situ formation, at $a\sim500$~AU there is typically insufficient time and material to build an Earth mass planet \cite{Kenyon,Heller,Ostriker}.  The prospect of a planet forming near Uranus and Neptune before being scattered to its present orbit is low since in order to fall into a stable orbit the planet would need to be appropriately influenced by a passing star (or another mechanism) \cite{Batygin-PR,Li}. The probability of capturing a free floating planet (FFP) is estimated to be similarly improbable, with estimates differing by orders of magnitude depending on assumptions \cite{Li,Parker,Mustill}.
 
We will argue that  while there is a low probability of capturing an Earth mass PBH, it is no more improbable than capturing an FFP of similar mass. The Solar System capture rate can be expressed as follows
\begin{equation}
\Gamma = \langle \sigma n v \rangle = \int n_0 F(v+v_{\odot}) ~\frac{{\rm d} \sigma}{{\rm d}v} ~v \mathrm{d}v~,
\label{eq:capturerate}
\end{equation}
where $F(v)$ and $n_0$ are the velocity distribution and ambient density of the objects to be captured, ${\rm d}\sigma/{\rm d}v$ is the differential capture cross-section and $v_{\odot,r}$ is the velocity of the Sun with respect to the rest frame of the objects to be captured. 

The differential cross-section (identical for PBHs and free floaters) is significantly suppressed for relative velocities larger than $0.25\mathrm{km/s}$ \cite{Goulinski}, which is much smaller than other velocities in the calculation. As a result, the velocity dispersions in the integrand can be approximated by the zero-order value $F(v_{\odot})$. This allows us to cancel the common factor of $\mathrm{d}\sigma/\mathrm{d}v$ in the ratio of PBH to FFP capture rate, which is then well approximated by
\beq
\frac{\Gamma_{\BH}}{\Gamma_{\FFP}} 
 \simeq \frac{n_{\BH}}{n_{\FFP}} \frac{F_{\PBH} (v_{\odot,\PBH})}{F_{\FFP}(v_{\odot,\FFP})}.
\eeq
We assume that the PBH velocity distribution is the same as the DM velocity distribution given by the Standard Halo Model \cite{Drukier:1986tm}, with $v_{\odot,\DM} = 220$km/s and velocity dispersion $\sigma_{\rm PBH} = v_\odot/\sqrt{2}$. The local density of PBHs is related to the local DM density ($\rho_{\rm DM} = 0.4\;\mathrm{GeV}/\mathrm{cm}^3$) and the fact that PBHs comprise a fraction  $f_{\rm PBH}$ of this local density:
\begin{equation}
n_{\BH} = f_{\PBH}\left(\frac{\rho_{\DM}}{M_{\BH}}\right) \sim 35 \mathrm{pc}^{-3} \left(\frac{f_{\BH}}{0.05}\right) \left(\frac{5 M_{\oplus}}{M_{\BH}}\right).
\notag
\end{equation}
As for the FFPs, there are two possibilities: While the Solar System could capture a planet when inside a star-forming region, for which the FFP density may be as high as $200\mathrm{pc}^{-3}$ \cite{Goulinski}, such stellar nurseries are highly disruptive environments. Hence a planet captured in this manner is quite likely subsequently stripped by interactions with nearby stars \cite{Li,Heller,Ostriker}. Instead we consider capture in the field, away from star forming regions. In the field, the FFP density (which we take to the be similar to the local star density) is much lower: $n_{\mathrm{FFP}}\sim0.2\mathrm{pc}^{-3}$. However, the available time for capture is much longer and the survival probability of a captured object is effectively unity. We assume the FFP velocity dispersion is inherited from the stars in the thin disk: $\sigma_* \sim 40 \mathrm{km/s}$ \cite{Goulinski}. Remarkably, with these parameters we arrive at:
\beq
\frac{\Gamma_{\BH}}{\Gamma_{\FFP}} 
& \sim 1\times \left(\frac{0.2\mathrm{pc}^{-3}}{n_{\FFP}}\right) \left(\frac{40 \mathrm{km/s}}{\sigma_{\FFP}}\right)^3 \left(\frac{f_{\BH}}{0.05}\right) \left(\frac{5 M_{\oplus}}{M_{\BH}}\right).
\notag
\eeq
We find that the rates are comparable and thus conclude that the probability that an FFP is gravitationally captured by the Solar System in ambient space is roughly comparable to capturing a $5M_\oplus$ PBH with $f_{\rm PBH}\sim0.05$. Therefore, if one is willing to entertain that the possibility that the TNO orbits indicate a captured planet, it is plausible that the gravitational source in the outer Solar System could instead be a PBH (once we establish evidence for such a PBH population).

Finally, we note that gravitational capture normally occurs due to multi-body interactions or drag through the local environment which leads to energy dissipation. Interestingly, capture of an object with an extended halo presents a new mechanism for dissipation since DM particles will be shed during the encounter. This possibly improves the chance of capture, but understanding the complicated dynamics would require a dedicated study.


\vspace{3mm}
{\bf 4.~Dark Matter.}~OGLE  \cite{Niikura:2019kqi} indicates  $f_{\PBH}\ll 1$. If the rest is taken up by the DM component, PBHs accrete dense DM microhalos. This is fortunate since DM annihilation provide a potential detection route and in the absence of DM it would be likely impossible to detect an $M_\oplus$ PBH in the Solar System.
To describe the DM profile around the PBH one needs to consider its initial configuration and subsequent evolution. We will discuss several characteristic scales that control the properties of the halo: the influence radius $r_{\rm in}$, truncation distances $r_t$ due to striping, and (if applicable) the radius at which DM annihilations shape the halo $r_{\rm max}$.

The radius of influence $r_{\rm in}$ corresponds to the region in which the PBH dominates the local gravitational potential (assuming the uniform background density approximation) and effectively appears as an $\mathcal{O}(1)$ density perturbation. As a result, $r_{\rm in}$ corresponds to the radius that contains DM mass equal to the PBH mass:
        \begin{equation}
            M_{\BH} = \frac{4\pi}{3} \rho(t) r_{\rm in}^3(t)~.
            \label{influ}
        \end{equation}
The density profile of a halo is relatively simple if the DM kinetic energy can be neglected relative to its potential energy.\footnote{When the kinetic energy cannot be neglected, an inner density plateau occurs associated to the DM free streaming scale  \cite{Eroshenko,Boucenna:2017ghj}.} This is typically the case for $M_{\PBH}\gtrsim M_\oplus$ for DM with mass $m\gtrsim100$ GeV \cite{Adamek:2019gns}, in which case the DM profile is of the form \cite{Bertschinger} (this profile is typical of self-similar secondary infall): 
\begin{equation}
\rho(r)= \frac{\rho_{\rm eq}}{2}\left(\frac{r_{\rm eq}}{r}\right)^{9/4}~,
\label{eq:profile}\end{equation}
in terms of $r_{\rm eq}\equiv r_{\rm in}(t_{\rm eq})\sim220~{\rm AU}\times(M_\BH/5M_\oplus)^{1/3}$ the radius of influence at matter-radiation equality, and 
 $\rho_{\rm eq}\equiv\rho(t_{\rm eq})\simeq 2.1\times10^{-19}~{\rm g}/{\rm cm}^3$ the density of Universe at matter-radiation equality.

If the DM can annihilate, the inner DM density may be depleted implying a cross-section dependent region of constant density $\rho_{\rm max} = m/\langle \sigma v \rangle \tau$ within a radius $r_{\rm max} = r_{\rm eq}\left[\langle \sigma v \rangle \tau \rho_{\rm eq}/(2m)\right]^{4/9}$ where $\langle \sigma v \rangle$ is the thermally averaged annihilation cross-section and $\tau$ is the age of the universe \cite{Eroshenko}. For the DM models we consider in this work $r_{\rm max}$ is smaller than the Schwarzchild radius $r_\BH$ and  density plateau does not exist. This happens for
\beq
\langle \sigma v \rangle < 
1.4\times 10^{-54} \mathrm{cm^3/s} \left(\frac{m}{100~\mathrm{GeV}}\right) \left(\frac{M_{\rm BH}}{5M_{\oplus}}\right)^{3/2}.
\label{rmaxrbh}
\eeq 

Furthermore, the DM halo will not have indefinite extent. It will be truncated due to stripping by the Milky Way (MW), by encounters with passing stars, and by the Sun (for a PBH bound to the Solar System). 
Since we consider a captured PBH, it should have a similar (galactic) orbit to the Sun, thus it will not make any especially close approaches to the center of the Galaxy. This implies very little truncation due to the Galaxy, for an initial halo of $M_{\mathrm{initial}} \sim 100 M_\oplus$ then $r_{t, {\rm MW}} \sim 10 \mathrm{kpc}$. 

Tidal disruptions from meeting individual stars can be more significant. The tidal radius $r_{t,\star}$ is dominated by the closest approach to a star. Since the PBH is travelling near the peculiar velocity of the Sun (10km/s) with respect to the local rest, it has travelled around 100 kpc over $10^{10}$ years and has passed $\mathcal{O}(10^5)$ stars. The typical spacing of stars is 1~pc, and so the distance of closest approach is $r_*\sim\sqrt{10^{-5}}\mathrm{pc} = 650\mathrm{AU}$. The tidal stripping radius due to star encounters is then
        \begin{equation}
             r_{t,\star} \sim r_*\left(\frac{M_{\mathrm{initial}}}{2M_\odot}\right)^{\frac{1}{3}}\sim  34\mathrm{AU}
              \left(\frac{r_*}{650~\mathrm{AU}}\right)  \left(\frac{M_{\mathrm{initial}}}{100 M_\oplus}\right)^{\frac{1}{3}}.
\notag
        \end{equation}
        Interestingly,  $r_*$ is of the same order as the inferred P9 semi-major axis. The halo mass, obtained by integrating to $r_{t,\star}\sim34$ AU, for the profile in eq.~(\ref{eq:profile}) reveals that the total halo mass is typically $\mathcal{O}(50\%)$ of a PBH of mass $5M_\oplus\lesssim M_{\rm BH}\lesssim10M_\oplus$, and so any further tidal stripping is controlled by the PBH mass. Once the PBH settles into an orbit around the Sun, tidal radius cuts off the DM halo at
        \begin{equation}
            r_{t,\odot} \sim r_p \left(\frac{M_{\rm BH}}{2M_\odot}\right)^{\frac{1}{3}} \;\sim\;
            8\mathrm{AU} \left(\frac{r_p}{400\mathrm{AU}}\right)\left(\frac{M_{\rm BH}}{5 M_\oplus}\right)^{\frac{1}{3}},
\notag        \end{equation}
which contains DM mass of the order $\mathcal{O}(15\%)$ of the mass of the PBH.
  

\vspace{3mm}
    \textbf{5.~Annihilation Signals.} 
On its own, a PBH of mass $5M_{\oplus}$ has a Hawking temperature of $0.004\;\mathrm{K}$, making it colder than the CMB, and since it's radius is $r_\BH \sim 5\;\mathrm{cm}$, the power radiated by the PBH alone is minuscule. However, the DM halo around this PBH 
can, if annihilating, provide a powerful signal. Annihilations in the PBH halo at the position of P9 would make for a potential FERMI-LAT source.  Moreover, the whole PBH population contributes to gamma ray diffuse emissions \cite{Bertone:2005xz}. Indeed, for a non-negligible $f_\PBH$ and DM with a thermal  cross-section $\langle\sigma v\rangle_0\sim3\times10^{-26}{\rm cm}^3/{\rm s}$  (i.e.~classic WIMP DM), the other PBHs surrounded by DM give a diffuse gamma ray flux that is strongly excluded \cite{Boucenna:2017ghj,Adamek:2019gns}.

In what follows, we take a DM model that generates observable signals, but is not yet excluded, and consider a ``freeze-in'' DM candidate  \cite{Hall:2009bx}. In freeze-in scenarios the DM abundance is initially negligible and is subsequently generated by highly suppressed interactions between the Standard Model and the DM, controlled by a tiny coupling $\lambda\ll1$, leading to a relic density $\Omega_{\rm DM}\propto m Y_{\rm FI} \propto \lambda^2$. 

For a specific model we consider a DM particle $\chi$ with mass $m$ coupled to a mediator state $\phi$ with mass $M_\phi$ via a Lagrangian term $g \bar{\chi} \chi \phi$, and in addition we couple $\phi$ to some Standard Model operator with coupling $\lambda$.  The DM relic density generated by freeze-in is parametrically 
\beq
\Omega_{\rm DM}
&\simeq
 0.2 \left(\frac{m}{100~{\rm GeV}}\right)
\left(\frac{\lambda}{6\times10^{-12}}\right)^2
\left(\frac{10~{\rm TeV}}{M_\phi}\right)~.
\label{relic}
\eeq

We assume an annihilation cross-section to Standard Model particles of the form (cf.~Supplementary Material)
\beq
\langle\sigma v \rangle \simeq \frac{\lambda^2 g^2m^2}{\pi M_\phi^4}~,
\label{xc}
\eeq
with the characteristic values in eq.~(\ref{relic}) this implies a characteristic cross-section of order
\beq
\langle\sigma v \rangle_{\mathrm{ch}}\simeq 
1.3 \times 10^{-56} {\rm cm}^{3}/{\rm s} \times
\bigg(\frac{g}{10^{-2}}\bigg)^2.
\label{xx}
\eeq
For fixed $\lambda$, decreasing $g$ reduces the annihilation cross-section while maintaining the DM abundance.   More model details appear in the Supplementary Material and related literature, see e.g.~\cite{Holdom:1985ag,Dienes:1996zr,Gherghetta:2019coi,Chu:2011be,Chu:2013jja,Elahi:2014fsa,Heeba:2019jho,Foldenauer:2018zrz,Alves:2015pea}.
We stress, however, that we present freeze-in as an example of a potentially discoverable scenario that is not currently ruled out.

\vspace{3mm}
   {\bf 5.a.~Point Source Limits.}   The flux for this freeze-in scenario can be found from the DM annihilation rate
      \beq
      \Gamma= 4\pi \int r^2 {\rm d}r \left(\frac{\rho(r)}{m} \right)^2 \langle \sigma  v \rangle.
   \label{lum}   
      \eeq 
    We take $ \langle \sigma  v \rangle\simeq \langle \sigma  v \rangle_{\rm ch}$, then $r_\mathrm{max} < r_\BH$ from eq.~(\ref{rmaxrbh})  and thus the profile of eq.~(\ref{eq:profile}) is appropriate. Cutting off the integral at the tidal stripping radius $r_{t,\odot}$ we obtain
\beq
\Gamma = \sqrt{\frac{3 \rho_\mathrm{eq}}{8\pi G^3}} \frac{\langle \sigma v \rangle }{m^2} =  10^{20} \mathrm{s}^{-1} \left(\frac{\langle\sigma v\rangle_{\;\;\;}}{\langle\sigma v\rangle_{\mathrm{ch}}} \right)\left(\frac{100\mathrm{GeV}}{m} \right)^{2}.
   \label{rate2}
   \eeq
Note that the above result is independent of the PBH mass (this is not the case for $r_\mathrm{max} > r_\BH$). Given the annihilation rate, the flux $\Phi_\gamma$ and the energy flux $\Phi_E$ of photons for an observer on Earth is then
\beq
\Phi_\gamma = \frac{\kappa_1 \Gamma}{4\pi r_9^2}~,
\hspace{5mm}
{\rm and}
\hspace{5mm}
\Phi_E = \frac{\kappa_2  \Gamma m}{4\pi r_9^2}~,
\notag \eeq
where $\kappa_1$ is the average number of photons per DM annihilation within the observable band of the experiment, while $\kappa_2$ is the fraction of energy of the DM annihilation converted into photons within the observable energy band. Both $\kappa_1$ and $\kappa_2$ are dependent on the mass and branching annihilation channel of the DM candidate and can be determined for example from \cite{Cirelli:2010xx}. To obtain characteristic bounds we set $\kappa_1 \sim 10$ and $\kappa_2 \sim 1$. 

The smallest detectable photon flux in the 8~year FERMI-LAT source catalog \cite{Fermi-LAT:2019yla} was J2143.0-5501 with $\Phi_\gamma = 8.8\times10^{-12} \mathrm{photons/cm}^2/\mathrm{s}$  in the band 100 MeV to 100 GeV, while the smallest energy flux in the same band was $\Phi_E = 5.96\times10^{-13}\mathrm{erg/cm}^2$/s due to 4FGL J1014.6+6126. Using these fluxes as upper bounds we calculate the maximum DM annihilation cross-section allowed by the photon flux limit, taking the rate of eq.~(\ref{rate2}) for $r_9\simeq400$ AU we obtain the bound
\beq
\langle \sigma v \rangle &< 
5.1 \times 10^{-56} \mathrm{cm}^3/\mathrm{s} \left(\frac{m}{100 \mathrm{GeV}}\right)^2 
\hspace{8mm} (\gamma~{\rm flux}).
\notag \eeq
The energy flux bound implies a comparable limit 
\beq
\langle \sigma v \rangle &< 2.2 \times 10^{-55} \mathrm{cm}^3/\mathrm{s} \left(\frac{m}{100 \mathrm{GeV}}\right) 
\hspace{8mm} (E~{\rm flux}).
\notag \eeq
That the two approaches give similar bounds is a coincidence and for different mass DM, these bounds will differ. Moreover, for more massive DM candidates these constraints weaken as the photons from annihilations become too energetic and different instruments, such as CTA \cite{Doro:2012xx}, will play an important role.

We note that during the 8 years of data taking the source would have moved by several degrees, making source identification much harder. However, the DM annihilation signals may be easier to spot because the space-time correlation in the signal would enhance the sensitivity of the search.  As a result, it is necessary to run a dedicated study in order to determine if there are any candidates that match the P9 trajectory in the FERMI-LAT dataset, which we will return to in future work \cite{futurework}.

\vspace{3mm}{\bf 5.b.~Diffuse Photons Limit.} In addition to looking for the P9 PBH there is also an observational bound from the full population of PBHs, this has previously been  explored in \cite{Adamek:2019gns,Bertone:2005xz,Lacki:2010zf,Eroshenko,Boucenna:2017ghj} for  thermal cross-section  WIMP DM. 

Following \cite{Adamek:2019gns,Boucenna:2017ghj} we bound the diffuse gamma ray flux due to PBH $\Phi_\gamma$ by translating the limits on decaying DM of mass $m$ and decay rate $\Gamma_{\rm DM}$ with the identification 
\beq
\frac{{\rm d}\Phi_\gamma}{{\rm d}E}
~\propto~\frac{f(1-f^2)\Gamma}{M_{\rm BH}}=\frac{\Gamma_{\rm DM}}{m}~.
\eeq
For DM with mass  $10~{\rm GeV}\lesssim m \lesssim10^{6}~{\rm GeV}$ the observational limit is 
\beq
\Gamma_{\rm DM}\lesssim10^{-28}~{\rm s}^{-1}
\eeq
assuming 100\% decays to $\overline{b}b$, varying by only $\mathcal{O}(1)$ over the mass range  \cite{Cohen:2016uyg}. Requiring that the differential flux due to $\Gamma$ is less than the above limit for $f=0.05$ implies a bound on the rate of
\beq
\Gamma\lesssim3.7\times10^{23}~{\rm s}^{-1}
\left(\frac{M_{\rm BH}}{5M_\oplus}\right)\left(\frac{m}{100~{\rm GeV}}\right).
\eeq
Comparing with the rate in eq.~(\ref{rate2}) with the characteristic cross-section of $\langle \sigma v \rangle_{\rm ch}$, it is seen that the diffuse bound is readily satisfied in the DM models we consider.
 
 {\bf 6.~Exotica.}
Before closing, we note that alternative `exotic' compact astrophysical bodies may explain these observations such as DM microhalos (without PBH) e.g.~\cite{Diemand:2005vz,Berezinsky:2007qu,Green:2005fa,Zhang:2010cj,Blanco:2019eij}, bose stars \cite{Jetzer:1991jr}, or  DM stars \cite{Kouvaris:2015rea,Curtin:2019lhm}. OGLE cannot distinguish between PBH, exotic stars, and planets, however DM microhalos are unlikely to produce OGLE's lensing events. Another possibility is a sizeable DM halo could be shredded during the capture leading a toroidal DM mass distribution around the Sun at $\sim 500$ AU with total mass $\sim10M_\oplus$, this realises the secular approximation (phase space averaged) for a compact object, and is similar to the proposed toroidal baryonic distribution of \cite{Sefilian}. Notably, each of these scenarios implies  different experimental signatures, distinct from those of a rocky or gas planet.

\vspace{3mm}
 {\bf 7.~Conclusion.}
This letter highlights that anomalous orbits of TNOs  and OGLE's short microlensing events could have the same origin and explores the intriguing scenario that they both arise due to
a population of $5M_\oplus$ PBHs. While the principal search strategies for a planet is to employ optical \cite{Linder,Ginzburg} and infrared/microwave surveys \cite{Meisner}, the signals could be very different for a PBH (or another exotic object). Thus, the PBH hypothesis expands the required experimental program to search for the body responsible for TNO shepherding and motivates dedicated searches for moving sources in x-rays, gamma rays and other high energy cosmic rays.
Conversely, if  conventional searches fail to find Planet 9 and the evidence for TNO anomalies continues to grow, the PBH P9 hypothesis will become a  compelling explanation.

\vspace{1mm}


\onecolumngrid

\vspace{3mm}

 {\bf Acknowledgements.}
We thank Martin Bauer for comments on the draft.
We are grateful for the hospitality and support of the University of Oxford and the Simons Center for Geometry and Physics (Program: Geometry \& Physics of Hitchin Systems). 
Part of this work was performed at the Aspen Center for Physics (ACP), which is supported by National Science Foundation grant PHY-1607611; the participation of JS at the ACP was supported by the Simons Foundation.
JS is also very grateful for the support from the COFUND Fellowship.
JU gratefully acknowledges support from the National Science Foundation grant  DMS-1440140 while in residence at MSRI during Fall 2019.

\vspace{-1mm}
\subsection*{{\sc \large Supplementary Material}}
\twocolumngrid

\vspace{-3mm}
\section*{A.~Size of the PBH}
\vspace{-2mm}

\begin{figure}[h!]
\vspace{1mm}
\includegraphics[height=9.35cm]{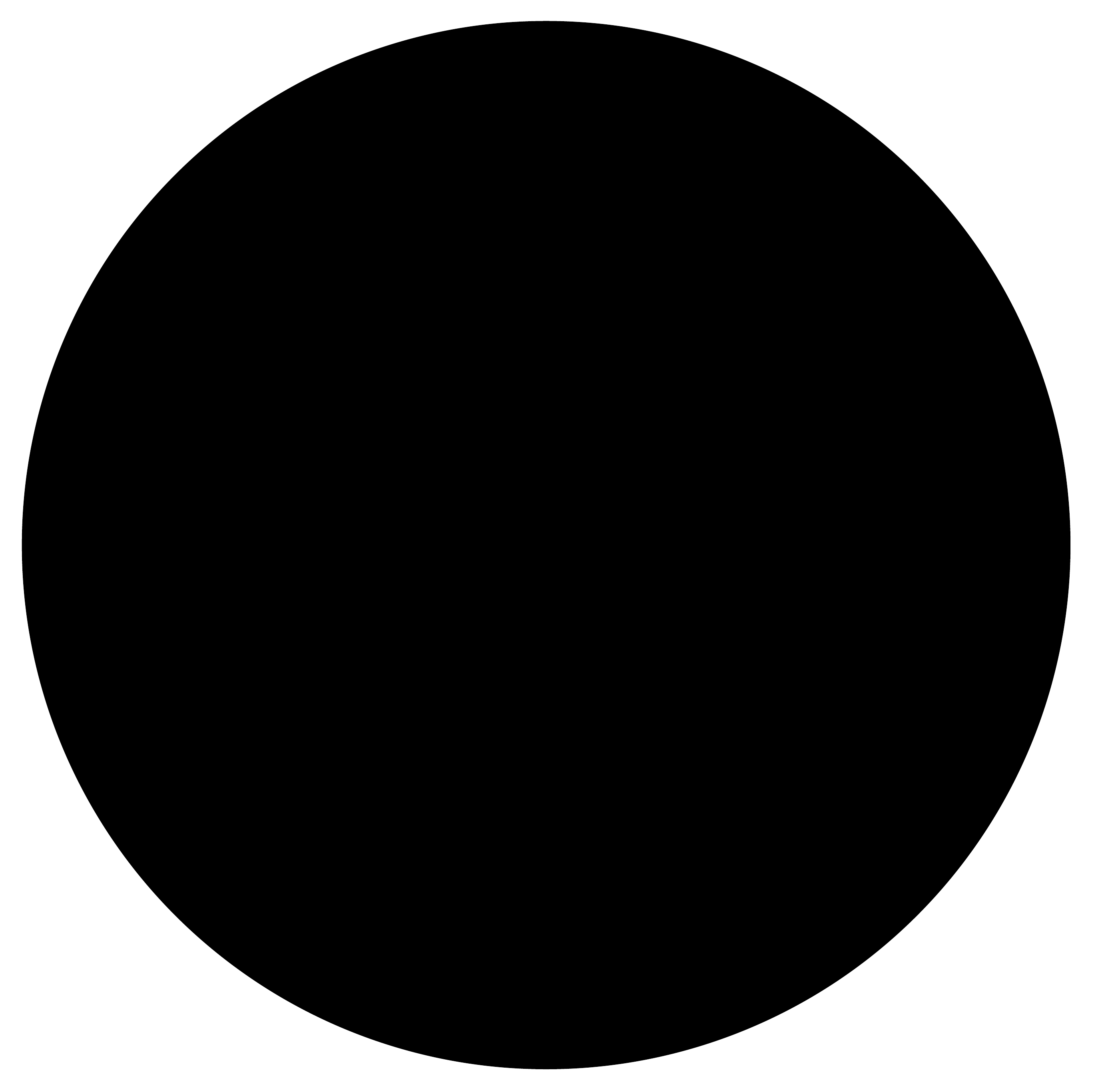}
	\caption{Exact scale (1:1) illustration of a $5M_\oplus$ PBH. Note that a  $10M_\oplus$ PBH is roughly the size of a ten pin bowling~ball.}\label{fig1}
\end{figure}

The Schwarzschild radius of a black hole is given by
\beq
r_{\rm BH}=\frac{2GM_{\rm BH}}{c^2}\simeq 4.5{\rm cm}\left(\frac{M_{\rm BH}}{5M_\oplus}\right)~.
\eeq
In Figure \ref{fig1} we provide an exact scale image of a $5M_\oplus$ PBH. The associated DM halo however extends to the stripping radius $r_{t,\odot}\sim8$AU, this would imply a DM halo which extends roughly the distance from Earth to Saturn (both in real life and relative to the image).

\onecolumngrid

\newpage
\twocolumngrid

\section*{B.~Dark Matter Freeze-in via a $Z'$}

To demonstrate a viable freeze-in model with potentially detectable indirect detection signals we consider a specific implementation in which the Standard Model is supplemented by two new ``hidden sector'' states, the DM particle $\chi$ with mass  $m$ and a mediator $Z'$ with mass $M_{Z'}$. We assume a sizeable interaction $gZ'\overline{\chi}\chi$ for $g\sim\mathcal{O}(10^{-2})$, with mass ordering $M_{Z'}>2m$, and that the hidden sector only couples to the Standard Model fermions $f$ (with mass $m_f$) via $\lambda Z'\overline{f}f$, which involves a tiny coupling  $\lambda\sim10^{-12}$. 
 
Here we relabel the mediator $Z'$ (from $\phi$ in the main text) and highlight that it can naturally arise from a  kinetic mixing portal \cite{Holdom:1985ag} involving a new U(1)$'$ gauge symmetry via the Lagrangian term $\Delta{\mathcal L}=\delta B_{\mu\nu}X^{\mu\nu}$
where $X$ is the $Z'$ field strength and $B$ is the hypercharge field strength, where $\delta$  parameterises the mixing induced by states  (with masses $M$ and $M'$)  charged under both U(1)$'$ and hypercharge $\delta\propto \frac{g_Yg_X}{32\pi^2}{\rm log}\left(M^2/M'{}^2\right)$. In motivated models this can lead to couplings appropriate for freeze-in \cite{Dienes:1996zr,Gherghetta:2019coi} and one can identify the freeze-in coupling $\lambda\simeq\delta g_{Y}$ where $g_Y$ are Standard Model fermion hypercharge couplings. For further details on $Z'$ freeze-in portals see e.g.~\cite{Chu:2013jja,Chu:2011be,Elahi:2014fsa,Heeba:2019jho,Foldenauer:2018zrz}.

The initial  $Z'$ and $\chi$ abundances are negligible but populations are generated via the portal interaction $\lambda Z'\overline{f}f$.
 The $Z'$ abundance evolves according to \cite{Hall:2009bx}
\beq
\dot n_{Z'}+3n_{Z'} H\simeq  \int & {\rm d}\Pi_{Z'}  {\rm d}\Pi_{\overline{f} }{\rm d}\Pi_{f}  (2\pi)^4\\
 &\delta^4(p_{Z'}-p_f-p_{\overline{f}})|\mathcal{M}|^2_{Z'\rightarrow \overline{f}f}f_{Z'}^{\rm eq}
\eeq
where $f_{Z'}^{\rm eq}={\rm exp}(-E_\chi/T)$ is the equilibrium distribution. 

The generated abundance of $Z'$ is  \cite{Hall:2009bx}
\beq
Y_{Z'}=\frac{135}{8\pi^3g_*^{S}\sqrt{g_*^{\rho}}}\left(\frac{M_{\rm Pl}\Gamma_{Z'}}{M_{Z'}^2}\right)
\eeq
where $\Gamma_{Z'}$ is the partial width involving bath states. 

Since we take $M_{Z'}>2m$ and $g\sim\mathcal{O}(10^{-2})$ any $Z'$ produced decays promptly to DM pairs and thus the resulting DM relic density is $Y_\chi\approx2Y_{Z'}$ for $\lambda\ll g$. This can be related to the observed quantity
\beq
\Omega_{\rm DM}&\simeq
 0.2 \left(\frac{m}{100~{\rm GeV}}\right)
\left(\frac{\lambda}{6\times10^{-12}}\right)^2
\left(\frac{10~{\rm TeV}}{M_{Z'}}\right)~,
\eeq 
where we have parameterised $\Gamma_{Z'}\simeq \lambda^2M_{Z'}/8\pi$. 

The feeble coupling $\lambda\ll1$ is characteristic of freeze-in models and is actually required to satisfy the consistency condition $Y_{\rm DM}(T)\ll Y_{\rm DM}^{\rm eq}$(T). If $\lambda$ is too large the hidden sector enters equilibrium ($Y_{\rm DM}\sim Y_{\rm DM}^{\rm eq}$) and then subsequently freeze-out dynamics sets the DM relic density.
However, since $\lambda$ is very small, direct and indirect detection is typically challenging. For the $Z'$ model the s-wave annihilation cross-section for $m_f\ll m<M_{Z'}$ is \cite{Alves:2015pea}
\beq
\sigma_0\simeq\sum_f \frac{N_c \lambda^2}{2\pi }
 \left(\frac{2g_{\chi v}^2   m^2}{(M_{Z'}^2-4m^2)^2 } + g_{\chi a}^2 \frac{m_f^2}{M_{Z'}^4}\right),
\eeq
where the sum runs over Standard Model fermions and $g_{\chi a}$ and $g_{\chi v}$ are the axial and vector couplings to $\chi$. Then eq.~(\ref{xc}) is obtained from the above in the limit $ m\ll M_{Z'}$. In the main text we highlight that if $f_{\rm PBH}$ is relatively large, indirect detection may be a viable route to search for this class of models in certain parameter ranges.

\onecolumngrid


\vspace{3mm}
\begin{thebibliography}{}


\twocolumngrid


\bibitem{Brown04}
M.~E.~Brown, C.~Trujillo, and D.~Rabinowitz,
{\em Discovery of a candidate inner Oort cloud planetoid}, 
ApJ 617.1 (2004): 645. [\href{https://arxiv.org/abs/astro-ph/0404456}{astro-ph/0404456}].

 \bibitem{Trujillo}
C.~Trujillo, and S.~S.~Sheppard, {\em A Sedna-like body with a perihelion of 80 astronomical units}, Nature 507.7493 (2014): 471.

\bibitem{Batygin}
K.~Batygin, and M.~E.~Brown, {\em Evidence for a distant giant planet in the solar system}, AJ 151.2 (2016): 22. [\href{https://arxiv.org/abs/1601.05438}{1601.05438}].

\bibitem{Batygin-PR}
K.~Batygin,  F.~C.~Adams, M.~E.~Brown, J.~C.~Becker, {\em The planet nine hypothesis},~Phys. Rep.~(2019). [\href{https://arxiv.org/abs/1902.10103}{1902.10103}].


 \bibitem{Mroz} P.~Mr{\'o}z, A.~Udalski, J.~Skowron,  {\em et al.} 
{\em No large population of unbound or wide-orbit Jupiter-mass planets}
 Nature 548 (2017) 183 [\href{https://arxiv.org/abs/1707.07634}{1707.07634}].

\bibitem{Niikura:2019kqi}
  H.~Niikura, {\em et al.}
{\em Constraints on Earth-mass primordial black holes from OGLE 5-year microlensing events,}
  Phys.\ Rev.\ D {\bf 99} (2019) no.8,  083503
 [\href{https://arxiv.org/abs/1901.07120}{1901.07120}].
      
  \bibitem{OGLE2}   P.~Mr{\'o}z,   {\em et al.}  [The OGLE Collaboration] {\em Two new free-floating or wide-orbit planets from microlensing}, A\&A, 622, A201  (2019) [\href{https://arxiv.org/abs/1811.00441}{1811.00441}].

  

\bibitem{Sheppard}
S.~Sheppard, {\em et al.}~{\em A New High Perihelion Inner Oort Cloud Object},  ApJ 157.4 (2019). [\href{https://arxiv.org/abs/1810.00013}{1810.00013}].

\bibitem{Gomes}
R.~S.~Gomes, J.~S.~Soares, and R.~Brasser, {\em The observation of large semi-major axis Centaurs: Testing for the signature of a planetary-mass solar companion}, Icarus 258 (2015): 37-49.

\bibitem{Gladman}
Gladman, B., {\em et al.} {\em Discovery of the first retrograde transneptunian object}, ApJ 697.2 (2009): L91.
 
 \bibitem{Chen}
Y.~T.~Chen, {\em et al.}~{\em Discovery of a new retrograde trans-neptunian object: hint of a common orbital plane for low semimajor axis, high-inclination TNOs and centaurs}, ApJ 827.2 (2016): L24. [\href{https://arxiv.org/abs/1608.01808}{1608.01808}].

\bibitem{Brown}
M.~E.~Brown,
{\em Observational bias and the clustering of distant eccentric Kuiper belt objects}, AJ 154.2 (2017): 65.
[\href{https://arxiv.org/abs/1706.04175}{1706.04175}].

\bibitem{Levison}
H.~Levison, {\em et al.} {\em Origin of the structure of the Kuiper belt during a dynamical instability in the orbits of Uranus and Neptune}, Icarus 196.1 (2008): 258-273.
[\href{https://arxiv.org/abs/0712.0553}{0712.0553}].

\bibitem{Nesvorny}
D.~Nesvorny, {\em Evidence for slow migration of Neptune from the inclination distribution of Kuiper belt objects.} ApJ 150.3 (2015): 73.
[\href{https://arxiv.org/abs/1504.06021}{1504.06021}].

\bibitem{Hawking:1971ei}
  S.~Hawking,
{\em Gravitationally collapsed objects of very low mass,}
  MNRAS {\bf 152} (1971) 75.

\bibitem{Carr:1974nx}
  B.~J.~Carr and S.~W.~Hawking,
   {\em Black holes in the early Universe,}
  MNRAS {\bf 168} (1974) 399.
  
\bibitem{Tada:2019amh}
  Y.~Tada and S.~Yokoyama,
  {\em Primordial black hole tower: Dark matter, earth-mass, and LIGO black holes,}
  Phys.\ Rev.\ D {\bf 100} (2019) no.2,  023537
  [\href{https://arxiv.org/abs/1904.10298}{1904.10298}].
  
  \bibitem{Fu:2019ttf}
  C.~Fu, P.~Wu and H.~Yu,
  {\em Primordial Black Holes from Inflation with Nonminimal Derivative Coupling,}
  [\href{https://arxiv.org/abs/1907.05042}{1907.05042}].
  
  \bibitem{Carr:2019kxo}
  B.~Carr, S.~Clesse, J.~Garcia-Bellido and F.~Kuhnel,
  {\em Cosmic Conundra Explained by Thermal History and Primordial Black Holes,}
  [\href{https://arxiv.org/abs/1906.08217}{1906.08217}].
  
  \bibitem{Jedamzik:1996mr}
  K.~Jedamzik,
  {\em Primordial black hole formation during the QCD epoch,}
  Phys.\ Rev.\ D {\bf 55} (1997) 5871
  [\href{https://arxiv.org/abs/astro-ph/9605152}{astro-ph/9605152}].

  \bibitem{Bertone:2005xz}
  G.~Bertone, A.~R.~Zentner and J.~Silk,
  {\em A new signature of DM annihilations: gamma-rays from intermediate-mass black holes,}
  Phys.\ Rev.\ D {\bf 72} (2005) 103517
  [\href{https://arxiv.org/abs/astro-ph/0509565}{astro-ph/0509565}].
  
  \bibitem{Lacki:2010zf}
  B.~C.~Lacki and J.~F.~Beacom,
  {\em Primordial Black Holes as DM: Almost All or Almost Nothing,}
  ApJ  {\bf 720} (2010) L67
  [\href{https://arxiv.org/abs/1003.3466}{1003.3466}].
  
    \bibitem{Eroshenko}
  Y.~N.~Eroshenko,
  {\em DM density spikes around primordial black holes,}
  Astron.\ Lett.\  {\bf 42} (2016) no.6,  347
   [Pisma Astron.\ Zh.\  {\bf 42} (2016) no.6,  359]
  [\href{https://arxiv.org/abs/1607.00612}{1607.00612}].
  
  \bibitem{Boucenna:2017ghj}
  S.~M.~Boucenna, F.~Kuhnel, T.~Ohlsson and L.~Visinelli,
  {\em Novel Constraints on Mixed Dark-Matter Scenarios of Primordial Black Holes and WIMPs,}
  JCAP {\bf 1807} (2018) no.07,  003
  [\href{https://arxiv.org/abs/1712.06383}{1712.06383}].
 
 \bibitem{Adamek:2019gns}
  J.~Adamek, C.~T.~Byrnes, M.~Gosenca and S.~Hotchkiss,
  {\em WIMPs and stellar-mass primordial black holes are incompatible,}
  Phys.\ Rev.\ D {\bf 100} (2019) no.2,  023506
  [\href{https://arxiv.org/abs/1901.08528}{1901.08528}].



   \bibitem{Kenyon} S.J.~Kenyon, B.C.~Bromley, {\em Making planet nine: Pebble accretion at 250-750 AU in a gravitationally unstable ring}, ApJ. 825, 33 (2016).
[\href{https://arxiv.org/abs/1603.08008}{1603.08008}].

  \bibitem{Heller} C.~H.~Heller, {\em Encounters with Protostellar Disks. II. Disruption and Binary Formation}, ApJ
  455, 252 (1995).

\bibitem{Ostriker} E.~C.~Ostriker, {\em Capture and Induced Disk Accretion in Young Star Encounters},  ApJ, 424, 292 (1994).

            \bibitem{Li} G.~Li, and F.~C.~Adams {\em Interaction cross-sections and Survival Rates for Proposed Solar System Member Planet Nine} ApJ 823.1 (2016): L3. [\href{https://arxiv.org/abs/1602.08496}{1602.08496}].

\bibitem{Mustill}  A.~Mustill, S.~Raymond, and M.~Davies, {\em Is there an exoplanet in the Solar system?} MNRAS Lett.\
 460.1 (2016): L109-L113. [\href{https://arxiv.org/abs/1603.07247}{1603.07247}].

\bibitem{Parker} 
 R.~Parker, T.~Lichtenberg, and S.~Quanz, {\em Was Planet 9 captured in the Sun's natal star-forming region}, 
  MNRAS Lett.\ 472.1 (2017): L75-L79. [\href{https://arxiv.org/abs/1709.00418}{1709.00418}].
  
  \bibitem{Goulinski}
  N.~Goulinski, and E.~Ribak, {\em Capture of free-floating planets by planetary systems.} 
  MNRAS 473.2 (2017): 1589-1595. 
  [\href{https://arxiv.org/abs/1705.10332}{1705.10332}].

\bibitem{Drukier:1986tm}
  A.~K.~Drukier, K.~Freese and D.~N.~Spergel,
  {\em Detecting Cold DM Candidates,}
  Phys.\ Rev.\ D {\bf 33} (1986) 3495.
      

  \bibitem{Bertschinger} E.~Bertschinger,  {\em Self-similar secondary infall and accretion in an Einstein-de Sitter universe}, ApJ, Suppl.\ S.\, 58, 39 (1985).

    
    \bibitem{Hall:2009bx}
  L.~J.~Hall, K.~Jedamzik, J.~March-Russell and S.~M.~West,
    {\em Freeze-In Production of FIMP DM,}
  JHEP {\bf 1003} (2010) 080
           [\href{https://arxiv.org/abs/0911.1120}{0911.1120}].
   
  
\bibitem{Holdom:1985ag}
  B.~Holdom,
  {\em Two U(1)'s and Epsilon Charge Shifts,}
  Phys.\ Lett.\  {\bf 166B} (1986) 196.
  
    \bibitem{Dienes:1996zr}
  K.~R.~Dienes, C.~F.~Kolda and J.~March-Russell,
   {\em Kinetic mixing and the supersymmetric gauge hierarchy,}
  Nucl.\ Phys.\ B {\bf 492} (1997) 104
  [\href{https://arxiv.org/abs/hep-ph/9610479}{hep-ph/9610479}].
  
  \bibitem{Gherghetta:2019coi}
  T.~Gherghetta, J.~Kersten, K.~Olive and M.~Pospelov,
  {\em The Price of Tiny Kinetic Mixing,}
     [\href{https://arxiv.org/abs/1909.00696}{1909.00696}].
  
    \bibitem{Chu:2011be}
  X.~Chu, T.~Hambye and M.~H.~G.~Tytgat,
  {\em The Four Basic Ways of Creating DM Through a Portal,}
  JCAP {\bf 1205} (2012) 034
   [\href{https://arxiv.org/abs/1112.0493}{1112.0493}].
  
\bibitem{Chu:2013jja}
  X.~Chu, Y.~Mambrini, J.~Quevillon and B.~Zaldivar,
  {\em Thermal and non-thermal production of DM via Z'-portal(s),}
  JCAP {\bf 1401} (2014) 034
 [\href{https://arxiv.org/abs/1306.4677}{1306.4677}].
  
    \bibitem{Elahi:2014fsa}
  F.~Elahi, C.~Kolda and J.~Unwin,
  {\em UltraViolet Freeze-in,}
  JHEP {\bf 1503} (2015) 048
     [\href{https://arxiv.org/abs/1410.6157}{1410.6157}].
  
  \bibitem{Heeba:2019jho}
  S.~Heeba and F.~Kahlhoefer,
    {\em Probing the freeze-in mechanism in dark matter models with $U(1)^\prime$ gauge extensions,}
       [\href{https://arxiv.org/abs/1908.09834}{1908.09834}].

\bibitem{Foldenauer:2018zrz}
P.~Foldenauer,
{\em Light dark matter in a gauged $U(1)_{L_\mu-L_\tau}$ model}
  Phys.\ Rev.\ D {\bf 99} (2019) 035007
 [\href{https://arxiv.org/abs/1808.03647}{1808.03647}].
  
     \bibitem{Alves:2015pea}
  A.~Alves, A.~Berlin, S.~Profumo and F.~S.~Queiroz,
 {\em DM Complementarity and the Z$^\prime$ Portal,}
  Phys.\ Rev.\ D {\bf 92} (2015) no.8,  083004
  [\href{https://arxiv.org/abs/1501.03490}{1501.03490}].
 
\bibitem{Cirelli:2010xx}
  M.~Cirelli {\it et al.},
  {\em PPPC 4 DM ID: A Poor Particle Physicist Cookbook for Dark Matter Indirect Detection,}
  JCAP {\bf 1103} (2011) 051
  [\href{https://arxiv.org/abs/1012.4515}{1012.4515}].
  
\bibitem{Fermi-LAT:2019yla}
  [Fermi-LAT Collaboration],
  {\em Fermi Large Area Telescope Fourth Source Catalog,}
  [\href{https://arxiv.org/abs/1902.10045}{1902.10045}].
  
\bibitem{Doro:2012xx}
  M.~Doro {\it et al.} [CTA Consortium],
  {\em Dark Matter and Fundamental Physics with the Cherenkov Telescope Array,}
  Astropart.\ Phys.\  {\bf 43} (2013) 189
    [\href{https://arxiv.org/abs/1208.5356}{1208.5356}].
  
    \bibitem{futurework}  
J.~Scholtz and J.~Unwin, work in progress.
    

\bibitem{Cohen:2016uyg} T.~Cohen, {\em et. al.}
{\em $\gamma$ -ray Constraints on Decaying Dark Matter and Implications for IceCube,}
  Phys.\ Rev.\ Lett.\ {\bf 119} (2017) no.2, 021102
    [\href{https://arxiv.org/abs/1612.05638}{1612.05638}].

\bibitem{Diemand:2005vz}
  J.~Diemand, B.~Moore and J.~Stadel,
  {\em Earth-mass dark-matter haloes as the first structures in the early Universe,}
  Nature {\bf 433} (2005) 389
  [\href{https://arxiv.org/abs/astro-ph/0501589}{astro-ph/0501589}].
  
  \bibitem{Berezinsky:2007qu}
  V.~Berezinsky, V.~Dokuchaev and Y.~Eroshenko,
  {\em Remnants of dark matter clumps,}
  Phys.\ Rev.\ D {\bf 77} (2008) 083519
  [\href{https://arxiv.org/abs/0712.3499}{0712.3499}].
  
  \bibitem{Green:2005fa}
  A.~M.~Green, S.~Hofmann and D.~J.~Schwarz,
  {\em The First wimpy halos,}
  JCAP {\bf 0508} (2005) 003
  [\href{https://arxiv.org/abs/astro-ph/0503387}{astro-ph/0503387}].
  
  \bibitem{Zhang:2010cj}
  D.~Zhang,
  {\em Impact of Primordial Ultracompact Minihaloes on the Intergalactic Medium and First Structure Formation,}
  MNRAS {\bf 418} (2011) 1850
    [\href{https://arxiv.org/abs/1011.1935}{1011.1935}].
  
\bibitem{Blanco:2019eij}
  C.~Blanco, M.~S.~Delos, A.~L.~Erickcek and D.~Hooper,
  {\em Annihilation Signatures of Hidden Sector Dark Matter Within Early-Forming Microhalos,}
  [\href{https://arxiv.org/abs/1906.00010}{1906.00010}].

\bibitem{Jetzer:1991jr}
  P.~Jetzer,
  {\em Boson stars,}
  Phys.\ Rept.\  {\bf 220} (1992) 163.

\bibitem{Kouvaris:2015rea}
  C.~Kouvaris and N.~G.~Nielsen,
  {\em Asymmetric Dark Matter Stars,}
  Phys.\ Rev.\ D {\bf 92} (2015) no.6,  063526
  [\href{https://arxiv.org/abs/1507.00959}{1507.00959}].

\bibitem{Curtin:2019lhm}
  D.~Curtin and J.~Setford,
  {\em How To Discover Mirror Stars,}
    [\href{https://arxiv.org/abs/1909.04071}{1909.04071}].
  
   \bibitem{Sefilian}
A.~Sefilian and J.~Touma, {\em Shepherding in a Self-gravitating Disk of Trans-Neptunian Objects},  2019,  AJ 157, 59
[\href{https://arxiv.org/abs/1804.06859}{1804.06859}].

\bibitem{Linder}
E.~F.~Linder, and C.~Mordasini, {\em Evolution and magnitudes of candidate Planet Nine}, A\&A 589 (2016): A134. 
[\href{https://arxiv.org/abs/1602.07465}{1602.07465}].

\bibitem{Ginzburg}
S.~Ginzburg, R.~Sari, and A.~Loeb, {\em Blackbody Radiation from Isolated Neptunes},   ApJ  822.1 (2016): L11.
 [\href{https://arxiv.org/abs/1603.02876}{1603.02876}].
  
\bibitem{Meisner}
%
A.~M.~Meisner, {\em et al.} {\em Searching for Planet Nine with coadded wise and neowise-reactivation images.}   ApJ 153.2 (2017): 65.
[\href{https://arxiv.org/abs/1611.00015}{1611.00015}].



\end{thebibliography}
\end{document}